\documentclass{article}%
\usepackage{amsmath}
\usepackage{amsfonts}
\usepackage{amsthm}
\usepackage{lineno,hyperref}
\usepackage{amssymb}
\usepackage{graphicx}
\usepackage{epstopdf}
\usepackage{caption}
\usepackage{subcaption}%
\usepackage{natbib}
\usepackage{authblk}
\newtheorem{hypothesis}{Hypothesis}
\newtheorem{theorem}{Theorem}
\newtheorem{proposition}[theorem]{Proposition}

\begin{document}
\vfill
{\centering%
\noindent\begin{tabular}{|l|}
%\centering
\hline
\textit{Sanz, L., Bravo de la Parra, R., 2020, Stochastic matrix metapopulation models}\\
\textit{with fast migration: Re-scaling survival to the fast scale. Ecological Modelling, 418, 108829.}\\
\textit{https://doi.org/10.1016/j.ecolmodel.2019.108829}\\
\hline
\end{tabular}
}%
\vfill

\title{Stochastic matrix metapopulation models with fast migration: re-scaling survival to the fast scale}

\author[1]{Luis Sanz\thanks{luis.sanz@upm.es}}
\author[2]{Rafael Bravo de la Parra \thanks{rafael.bravo@uah.es}}

\affil[1]{Depto. Matem\'aticas, E.T.S.I Industriales, Technical University of Madrid, Madrid, Spain.}
\affil[2]{U.D. Matem\'aticas, Universidad de Alcal\'a, Alcal\'a de Henares, Spain.}
\date{}

\maketitle

\begin{abstract}
In this work we address the analysis of discrete-time models of structured metapopulations subject to environmental stochasticity. Previous works on these models made use of the fact that migrations between the patches can be considered fast with respect to demography (maturation, survival, reproduction) in the population. It was assumed that, within each time step of the model, there are many fast migration steps followed by one slow demographic event. This assumption allowed one to apply approximate reduction techniques that eased the model analysis. It is however a questionable issue in some cases since, in particular, individuals can die at any moment of the time step. We propose new non-equivalent models in which we re-scale survival to consider its effect on the fast scale.
We propose a more general formulation of the approximate reduction techniques so that they also apply to the proposed new models. We prove that the main asymptotic elements in this kind of stochastic models, the Stochastic Growth Rate (SGR)\footnote{Stochastic Growth Rate (SGR). Scaled Logarithmic Variance (SLV)} and the Scaled Logarithmic Variance (SLV), can be related between the original and the reduced systems, so that the analysis of the latter allows us to ascertain the population fate in the first. Then we go on to considering some cases where we illustrate the reduction technique and show the differences between both modelling options. In some cases using one option
represents exponential growth, whereas the other yields extinction.
\end{abstract}

{\bf keywords:}metapopulation models, environmental stochasticity, time scales, approximate aggregation.

%\MSC[2010] 39A50 \sep 92D25
%\end{keyword}
%\end{frontmatter}

%\linenumbers

\section{Introduction}

\label{sec1} Mathematical models used in ecology, trying to mimic natural
systems complexity, are often described by a large number of variables
corresponding to various interacting organization levels. The use of reduction
and approximation techniques is a common approach in the analysis of the
proposed models. Among these techniques we could include the so-called
\textit{aggregation of variables} methods \citep{Auger08}. They are in
consonance with the important issue of up-scaling in the framework of
ecological hierarchy theory \citep{Lischke07}.

In this context, we consider stochastic linear discrete systems as population
dynamics models. We distinguish time scales to deal with the complexity of
these models. Our approach consists in assuming the existence of two different
processes acting together at different time scales. Both processes are
represented in matrix form, each of them in its associated time scale. We
choose as time unit of the common discrete model the characteristic one of the
slow process. We suppose that the slow time unit is approximately $k$ times
larger than the fast one and, further, that in this large interval the fast
process sequentially acts $k$ times followed by the slow process acting once.
Thus, the combined effect of both processes can be described by the product of
the slow-process matrix times the k-th power of the fast-process matrix.

The treated models incorporate environmental stochasticity, which refers to
unpredictable temporal fluctuations in environmental conditions. A number of
different environmental conditions are considered and their variation is
characterized by a sequence of random variables corresponding to the different
time steps of the discrete system \citep{Tulja90,Tulja97,Caswell01}. Each
environment is characterized by its corresponding matrix of vital rates. Given
certain hypotheses on the pattern of temporal variation and the vital rates in
each environment, the distribution of total population size is asymptotically
lognormal, with an expected value and a variance dependent on two parameters.
The first one is the \textit{stochastic growth rate} $\lambda_{\mathrm{S}}$
(SGR), which is the stochastic analogue of the dominant eigenvalue for
deterministic systems (we follow Cohen's definition for the SGR
\citep{Cohen79} because it allows one to directly compare the stochastic rate
of growth with its deterministic analogue, although some authors
\citep{Tulja90,Caswell01} define the SGR as the logarithmic rate of growth,
i.e., $\log\lambda_{\mathrm{S}}$). The second one is the \textit{scaled
logarithmic variance} (SLV), which characterizes the asymptotic deviation of
the population size from its mean value. The analytical derivation of the SGR
and SLV is not feasible in most situations, so it is necessary to resort to
computer simulations.

The reduction of two-scale discrete systems for populations subject to
environmental stochasticity was introduced in \cite{Sanz00}, where some
results relating the moments of the solutions of original and the reduced
systems were provided. In \cite{Sanz07} we studied how the SGR of the original
system can be approximated by the reduced system SGR for a finite number of
environments and a Leslie-type demography. Those results were extended in
\cite{Alonso09} to the case of an infinite number of different environmental
conditions and the relationship between the SLV of the original and the
reduced systems was established.

In this work, we focus on the analysis of discrete-time models of structured
metapopulations with environmental stochasticity and two time scales. Its
purpose is extending to the stochastic case the study performed in
\cite{Nguyen11} for deterministic models. In the aforementioned way of
separating the slow and the fast process one includes in the former all
mechanisms having to do with local demography (maturation, survival,
reproduction) and in the fast one the migration between patches. In
\cite{Nguyen11} it is argued that this way of separating the slow and the fast
process might be unrealistic in some ecological situations. Indeed, this
approach can be seen as assuming that individuals perform at first a series of
$k$ migration events followed by a demographic event in the arrival patch. The
assumption sounds realistic for reproduction when discrete models describe
populations with separated generations, but not so much for processes like
survival, since deaths can happen in any moment of the slow time step. To
reflect this point in the model we propose to alternatively include survival
in the fast process approximating its effect during the fast time unit, and we
refer to it as the \textit{re-scaling} of survival to the fast time scale.

The aim of this work is to adapt the method of re-scaling to the proposed
stochastic models and to compare the asymptotic behaviour of the models with
and without re-scaled survival. In order to do so, we use the aforementioned
reduction techniques, that need to be extended (see \ref{app2}) so that they
also apply to the new model with re-scaled survival.

The re-scaling procedure, described here for survival in the context of
structured metapopulations models, might be generalized to a much larger
setting. Nevertheless, we consider that it is more accessible to present it in
this particular yet still meaningful case.

The structure of the paper is as follows:

In Section \ref{sec21} we present two different metapopulation models for a
structured population subject to environmental stochasticity. The migration
among the different patches is assumed fast with respect to demography in the
first of these models. In the second one, part of demography, survival, is
re-scaled to the fast scale. With the help of the results in \ref{app2}, we
propose in Section \ref{sec22} a reduction technique of both models to
simplify their analysis and comparison.

The results of Section \ref{sec2} are applied in Section \ref{sec3}. In
\ref{sec31} we treat the case of an unstructured population in a multipatch
environment where, with the help of the reduction technique, we obtain
approximate closed expressions of the parameters defining the long term
behaviour of the solutions in both models. In \ref{sec32} some results for two
classes model are established. Finally, in \ref{sec33}, with the help of a
particular simple case, the relationships between both modeling approaches are discussed.

Following a discussion, two appendices are included. \ref{app1} briefly
introduces the basic form of matrix models with environmental stochasticity
and the result on the distribution of population size when the environmental
variation is a homogeneous Markov chain. \ref{app2} includes the presentation
of a general reducible linear discrete system with environmental
stochasticity, defined by means of sequences of matrices with limits suitable
for reduction, together with the construction of its reduced system and the
relationship between their corresponding asymptotic elements.

\section{Methods}

\label{sec2}

\subsection{Two-scale structured metapopulations models with environmental
stochasticity}

\label{sec21} In this section, we present a stochastic discrete population
model whose dynamics is driven by two processes, slow and fast, whose
corresponding characteristic time scales are very different from each other.
The population is considered structured into $q$ classes and inhabiting an
environment divided into $r$ patches. We first assume that the fast process
has to do with the movements of the individuals between patches whereas the
slow process encompasses all the demographic issues: births, deaths and
transitions between classes. In a second step, we will undertake the
re-scaling to the fast scale of the death process.

We represent the state of the population at time $t$ by vector
\[
\boldsymbol{x}(t):=(\boldsymbol{x}^{1}(t),\ldots,\boldsymbol{x}^{q}%
(t))^{\mathsf{T}}\in\mathbb{R}_{+}^{qr},
\]
where $\boldsymbol{x}^{i}(t):=(x^{i1}(t),\ldots,x^{ir}(t))\in\mathbb{R}%
_{+}^{r}$ and $x^{i\alpha}(t)$ denotes the population density of class $i$ in
patch $\alpha$.

We choose as the projection interval of our model the one corresponding to the
slow dynamics, i.e., the time elapsed between times $t$ and $t+1$, and we
denote it by $\Delta_{t}$.

We assume that the population can be subject to $n$ different environmental
conditions that we consider indexed by the set $\mathcal{I}=\left\{
1,...,n\right\}  $. The environmental variation is characterized by a sequence
of random variables $\tau_{t}$, $t=0,1,2,...$ defined on a certain probability
space $(\Omega,\mathcal{F},p)$ \citep{Billingsley12} over the state space
$\mathcal{I}$. For each realization $\omega\in\Omega$ of the process, the
population is subject to environmental conditions $\tau_{t+1}(\omega)$ between
times $t$ and $t+1$.

For each environment $\eta\in\mathcal{I},$ the slow process is defined by a
nonnegative projection matrix $\mathbf{D}_{\eta}=\left[  \boldsymbol{D}_{\eta
}^{ij}\right]  _{1\leq i,j\leq q}\in\mathbb{R}_{+}^{qr\times qr}$, divided
into blocks $\boldsymbol{D}_{\eta}^{ij}=\text{diag}\left(  d_{\eta}%
^{ij,\alpha}\right)  _{\alpha=1,\ldots,r}\in\mathbb{R}_{+}^{r\times r},$ where
$d_{\eta}^{ij,\alpha}$ represents the rate of individual's transition from
class $j$ to class $i$ in patch $\alpha$ during the slow time interval in
environment $\eta\in\mathcal{I}$.

Note that, for each environment $\eta$, matrix $\left[  d_{\eta}^{ij,\alpha
}\right]  _{1\leq i,j\leq q}\in\mathbb{R}_{+}^{q\times q}$ represents
demography in patch $\alpha$. This matrix does not appear as such in our
formulation due to the chosen ordering of the state variables.

Concerning the fast process, let $p_{\eta}^{i,\alpha\beta}$ represent the rate
of migration from patch $\beta$ to patch $\alpha$ for individuals of class
$i$. Therefore $\boldsymbol{P}_{\eta}^{i}=\left[  p_{\eta}^{i,\alpha\beta
}\right]  _{1\leq\alpha,\beta\leq r}\in\mathbb{R}_{+}^{r\times r}$ is a
column-stochastic matrix that we assume primitive, meaning the ability of
individuals initially present in every patch to eventually reach any other
patch. Finally, the matrix representing the fast process for the whole
population and each $\eta$ is $\mathbf{P}_{\eta}=\text{diag}\left(
\boldsymbol{P}_{\eta}^{i}\right)  _{i=1,\ldots,q}\in\mathbb{R}_{+}^{qr\times
qr}$

We now introduce the first way of modeling the fast-slow system. Since the
time step of the complete model is the one corresponding to the slow dynamics,
we need to approximate the effect of the fast process over this time interval.
In order to do so, we assume that the fast process acts $k$ times before the
slow process does, where $k$ might take a large value. Let $\tau_{t+1}$ be the
environmental process that selects the environment to which the population is
subjected during interval $\Delta_{t}$. Then, the complete system reads as
follows%
\begin{equation}
\boldsymbol{x}_{k}(t+1)=\mathbf{D}_{\tau_{t+1}}\left(  \mathbf{P}_{\tau_{t+1}%
}\right)  ^{k}\boldsymbol{x}_{k}(t), \label{modcomp2}%
\end{equation}
where $(\mathbf{A})^{k}$ denotes the $k$-power of matrix $\mathbf{A}$.

Model \eqref{modcomp2} can be interpreted as making individuals first perform
a series of $k$ dispersal events followed by the demographic process that
occurs in the patch of arrival. This assumption might be realistic in some
aspects, such as reproduction, because discrete models are mainly used for
species having offspring once every time unit. However, the situation is
different in the case of mortality since individuals may die at any time.

Therefore, we propose next a new version of model \eqref{modcomp2} in which
mortality acts at the fast time scale, and we call this re-scaling of survival
to the fast time scale.

Let $s_{\eta}^{j,\alpha}>0$ be the survival rate of class $j$ in patch
$\alpha$ and environment $\eta$. First, we factor every demographic
coefficient in order to make $s_{\eta}^{j,\alpha}$ appear explicitly, i.e., we
define $\tilde{d}_{\eta}^{ij,\alpha}$ through%
\[
d_{\eta}^{ij,\alpha}=\tilde{d}_{\eta}^{ij,\alpha}\,s_{\eta}^{j,\alpha}.
\]
Defining $\boldsymbol{\tilde{D}}_{\eta}^{ij}=\text{diag}\left(  \tilde
{d}_{\eta}^{ij,\alpha}\right)  _{\alpha=1,\ldots,r}$ and $\boldsymbol{S}%
_{\eta}^{j}=\text{diag}\left(  s_{\eta}^{j,\alpha}\right)  _{\alpha
=1,\ldots,r}$ we have
\[
\boldsymbol{D}_{\eta}^{ij}=\boldsymbol{\tilde{D}}_{\eta}^{ij}\,\boldsymbol{S}%
_{\eta}^{j}.
\]
Similarly, denoting $\mathbf{\tilde{D}}_{\eta}=\left[  \boldsymbol{\tilde{D}%
}_{\eta}^{ij}\right]  _{1\leq i,j\leq q}$ and $\mathbf{S}_{\eta}%
=\text{diag}\left(  \boldsymbol{S}_{\eta}^{j}\right)  _{j=1,\ldots,q}$ we
have
\[
\mathbf{D}_{\eta}=\mathbf{\tilde{D}}_{\eta}\,\mathbf{S}_{\eta}.
\]
Now, we make power $k$ appear explicitly in the survival rates by defining
matrices $\boldsymbol{S}_{k,\eta}^{j}=\text{diag}\left(  (s_{\eta}^{j,\alpha
})^{1/k}\right)  _{\alpha=1,\ldots,r},$ and $\mathbf{S}_{k,\eta}%
=\text{diag}(\boldsymbol{S}_{k,\eta}^{j})_{j=1,\ldots,q},$ that verify
$(\boldsymbol{S}_{k,\eta}^{j})^{k}=\boldsymbol{S}_{\eta}^{j}$ and
$(\mathbf{S}_{k,\eta})^{k}=\mathbf{S}_{\eta}$.

Finally we obtain the following expression for matrix $\mathbf{D}_{\eta}$
\[
\mathbf{D}_{\eta}=\mathbf{\tilde{D}}_{\eta}\,\left(  \mathbf{S}_{k,\eta
}\right)  ^{k}.
\]
This expression suggests the next complete model, not equivalent to model
\eqref{modcomp2}, in which it is considered that mortality acts as the fast
time scale:
\begin{equation}
\tilde{\boldsymbol{x}}_{k}(t+1)=\mathbf{\tilde{D}}_{\tau_{t+1}}\left(
\mathbf{S}_{k,\tau_{t+1}}\,\mathbf{P}_{\tau_{t+1}}\right)  ^{k}\tilde
{\boldsymbol{x}}_{k}(t) \label{modcomp3}%
\end{equation}
This model can be interpreted as follows: individuals first perform a series
of $k$ dispersal events in which in each of them we consider mortality by
taking into account the survival rate in the patch of arrival. This is
followed by the rest of the demographic process, that occurs at the slow time scale.

\subsection{Reduction of systems \eqref{modcomp2} and \eqref{modcomp3}}

\label{sec22} Following \ref{app2}, for system \eqref{modcomp2} we can take
\[
\boldsymbol{H}_{k,\eta}=\mathbf{D}_{\eta}\left(  \mathbf{P}_{\eta}\right)
^{k},
\]
and now we must express its limit when $k$ tends to infinity in the
appropriate form so that Hypotheses \ref{HA} and \ref{HB} are met.

The fact that matrix $\boldsymbol{P}_{\eta}^{i}$ is column-stochastic and
primitive implies that 1 is its dominant eigenvalue, the row vector
$\boldsymbol{\bar{1}}=(1,\ldots,1)\in\mathbb{R}^{r}$ is an associated left
eigenvector and there exists a unique positive right eigenvector
$\boldsymbol{v}_{\eta}^{i}\in\mathbb{R}^{r}$ such that $\boldsymbol{\bar{1}%
}\,\boldsymbol{v}_{\eta}^{i}=1$. The Perron--Frobenius theorem yields that
\[
\lim_{k\rightarrow\infty}\left(  \boldsymbol{P}_{\eta}^{i}\right)
^{k}=\boldsymbol{v}_{\eta}^{i}\,\boldsymbol{\bar{1}.}%
\]
Calling $\mathbf{V}_{\eta}:=\text{diag}\left(  \boldsymbol{v}_{\eta}%
^{i}\right)  _{i=1,\ldots,q}\in\mathbb{R}_{+}^{qr\times q}$ and $\mathbf{G}%
:=\text{diag}\left(  \boldsymbol{\bar{1}}\right)  _{i=1,\ldots,q}\in
\mathbb{R}_{+}^{q\times qr}$, we have that
\[
\lim_{k\rightarrow\infty}\left(  \mathbf{P}_{\eta}\right)  ^{k}=\mathbf{V}%
_{\eta}\,\mathbf{G}\ \text{ and so }\ \lim_{k\rightarrow\infty}\mathbf{D}%
_{\eta}\left(  \mathbf{P}_{\eta}\right)  ^{k}=\mathbf{D}_{\eta}\,\mathbf{V}%
_{\eta}\,\mathbf{G}.
\]
Thus, matrix $\mathbf{D}_{\eta}\,\mathbf{V}_{\eta}$ plays the role of matrix
$\boldsymbol{D}_{\eta}$ in \eqref{dfr}. We can now write the reduced system
\eqref{modagreg} for the global variables $\boldsymbol{y}(t)=\mathbf{G}%
\boldsymbol{x}(t)$, the total number of individuals in each class associated
to system \eqref{modcomp2}, in the following form:
\begin{equation}
\boldsymbol{y}(t+1)=\mathbf{G}\,\mathbf{D}_{\tau_{t+1}}\mathbf{V}_{\tau_{t+1}%
}\boldsymbol{y}(t)=\boldsymbol{\hat{H}}_{\tau_{t+1}}\boldsymbol{y}(t).
\label{modagreg2}%
\end{equation}

Let us now proceed to the reduction of system \eqref{modcomp3}, for which we
have, still following \ref{app2},
\[
\boldsymbol{H}_{k,\eta}=\mathbf{\tilde{D}}_{\eta}\left(  \mathbf{S}_{k,\eta
}\,\mathbf{P}_{\eta}\right)  ^{k}.
\]
To show that Hypotheses \ref{HA} and \ref{HB} also hold for system
\eqref{modcomp3}, we use Theorem A.1. in \cite{Nguyen11} to express in a
suitable form the limit of matrix $\boldsymbol{H}_{k,\eta}$ when $k$ tends to
infinity. For every $i=1,\ldots,q$ and $\eta\in\mathcal{I}$, let
$\boldsymbol{w}_{\eta}^{i}=\left(  \log(s_{\eta}^{i,1}),\ldots,\log(s_{\eta
}^{i,r})\right)  \in\mathbb{R}^{r}$ be a row vector and define the scalar
$\gamma_{\eta}^{i}=\exp\left(  \boldsymbol{w}_{\eta}^{i}\,\boldsymbol{v}%
_{\eta}^{i}\right)  $ then, \citep{Nguyen11},
\[
\lim_{k\rightarrow\infty}\left(  \boldsymbol{S}_{k,\eta}^{i}\,\boldsymbol{P}%
_{\eta}^{i}\right)  ^{k}=\gamma_{\eta}^{i}\boldsymbol{v}_{\eta}^{i}%
\,\boldsymbol{\bar{1}}.
\]
Calling $\mathbf{\tilde{V}}_{\eta}:=\text{diag}\left(  \gamma_{\eta}%
^{i}\boldsymbol{v}_{\eta}^{i}\right)  _{i=1,\ldots,q}$ it follows that
\[
\lim_{k\rightarrow\infty}\mathbf{\tilde{D}}_{\eta}\left(  \mathbf{S}_{k,\eta
}\,\mathbf{P}_{\eta}\right)  ^{k}=\mathbf{\tilde{D}}_{\eta}\,\mathbf{\tilde
{V}}_{\eta}\,\mathbf{G},
\]
and, therefore, the reduced system \eqref{modagreg} associated to system
\eqref{modcomp3} for the same global variables $\boldsymbol{\tilde{y}%
}(t)=\mathbf{G}\mathrm{\tilde{x}}(t)$, is
\begin{equation}
\boldsymbol{\tilde{y}}(t+1)=\mathbf{G}\,\mathbf{\tilde{D}}_{\tau_{t+1}}
\mathbf{\tilde{V}}_{\tau_{t+1}}\boldsymbol{\tilde{y}}(t)=\boldsymbol{\tilde
{H}}_{\tau_{t+1}}\boldsymbol{\tilde{y}}(t). \label{modagreg3}%
\end{equation}

The interest of the proposed reduction method is that it is possible to obtain
asymptotic results for systems \eqref{modcomp2} and \eqref{modcomp3} through
the analysis of the reduced systems \eqref{modagreg2} and \eqref{modagreg3}.

We need to impose some conditions so that the systems involved have indeed
good asymptotic properties. Theorem \ref{fust} in \ref{app1} presents
sufficient hypotheses so that the total population size of a matrix model with
environmental stochasticity is asymptotically lognormal and describable in
terms of a couple of constants, the stochastic growth rate (SGR) and the
scaled logarithmic variance (SLV). In \ref{app2} the reduction of matrix
models with environmental stochasticity is presented in a general setting. In
particular, it is proved that under suitable hypotheses for both the complete
and the reduced systems, the total population size is asymptotically
lognormal. Moreover, the SGR and the SLV of the complete system can be
approximated by those corresponding to the reduced system.

Assume that the following two conditions are met:

\begin{itemize}
\item[C1.] The environmental variation $\tau_{t}$ is a homogeneous irreducible
and aperiodic Markov chain, i.e., its matrix of transition probabilities is
primitive. We denote its (unique and positive) stationary probability
distribution by $\boldsymbol{\pi}=(\pi_{1},\ldots,\pi_{n})$.

\item[C2.] The sets $\left\{  \boldsymbol{\hat{H}}_{1},\ldots,\boldsymbol{\hat
{H}}_{n}\right\}  $ and $\left\{  \boldsymbol{\tilde{H}}_{1},\ldots
,\boldsymbol{\tilde{H}}_{n}\right\}  $ are ergodic, i.e., for each of them
there exists a positive integer $g$ such that any product of $g$ matrices
(with repetitions allowed) drawn from the set is a positive matrix. It is easy
to check that for each $\eta\in\mathcal{I},$ the incidence matrix of
$\boldsymbol{\tilde{H}}_{\eta}$ coincides with that of $\boldsymbol{\hat{H}%
}_{\eta}$ and, therefore, if one of the previous sets is ergodic so is the other.
\end{itemize}

Now we apply Proposition \ref{prop500} and Theorem \ref{th11} in \ref{app2}
and obtain that systems \eqref{modagreg2} and \eqref{modagreg3}, and for $k$
large enough also systems \eqref{modcomp2} and \eqref{modcomp3}, verify that
the total population size is asymptotically lognormal. Moreover, denoting
their corresponding SGR and SLV by $\hat{\lambda}_{\mathrm{S}}$,
$\tilde{\lambda}_{\mathrm{S}}$, $\hat{\lambda}_{\mathrm{S},k},$ $\tilde
{\lambda}_{\mathrm{S},k}$ and $\hat{\sigma}^{2}$, $\tilde{\sigma}^{2}$,
$\sigma_{k}^{2},$ $\tilde{\sigma}_{k}^{2}$ respectively we have
\begin{equation}%
\begin{array}
[c]{lll}%
\underset{k\rightarrow\infty}{\lim}\lambda_{\mathrm{S},k}=\hat{\lambda
}_{\mathrm{S}} & , & \underset{k\rightarrow\infty}{\lim}\sigma_{k}=\hat
{\sigma}\\
\underset{k\rightarrow\infty}{\lim}\tilde{\lambda}_{\mathrm{S},k}%
=\tilde{\lambda}_{\mathrm{S}} & , & \underset{k\rightarrow\infty}{\lim}%
\tilde{\sigma}_{k}=\tilde{\sigma}%
\end{array}
\label{app}%
\end{equation}
Thus, the lognormal asymptotic distribution for the complete system
\eqref{modcomp2} can be approximated through the constants $\hat{\lambda
}_{\mathrm{S}}$ and $\hat{\sigma}$ associated to the reduced system
\eqref{modagreg2}. The same happens to system \eqref{modcomp3} using the
constants $\tilde{\lambda}_{\mathrm{S}}$ and $\tilde{\sigma}$ associated to
\eqref{modagreg3}. Therefore, in order to establish a comparison between
systems \eqref{modcomp2} and \eqref{modcomp3}, we can compare the SGR and the
SLV of their corresponding simpler aggregated systems.

\section{Results}

\label{sec3} In this section we present some results comparing the two
modelling options represented by systems \eqref{modcomp2} and
\eqref{modcomp3}. Specifically, we carry out this comparison through their
respective SGRs and SLVs.

In most ecological models, the exact derivation of the SGR and the SLV is not
feasible. This is due to the fact that there is not an explicit expression for
the stationary distribution of the population structure. Therefore, it is
necessary to approximate those parameters by computer simulations or
appropriate perturbation techniques \citep{Tulja97}. The results presented in
Section \ref{sec22} justify using the simpler aggregated systems to carry out
these simulations. Moreover, there are particular cases in which our reduction
procedure transforms a complex model into a reduced one for which we can
obtain the SGR and the SLV exactly.

We proceed to introduce two such cases.

\subsection{Unstructured models}

\label{sec31} We first consider the models of Section \ref{sec21} when $q=1$,
i.e., there is no structure in the population.

We represent the state of the population at time $t$ by vector
\[
\boldsymbol{x}(t):=\left(  x^{1}(t),\ldots,x^{r}(t)\right)  \in\mathbb{R}%
_{+}^{r}\, ,
\]
where $x^{\alpha}(t)$ denotes the population density in patch $\alpha$. For
each environment $\eta\in\mathcal{I},$ the slow process is defined by the
nonnegative matrix $\mathbf{D}_{\eta}=\text{diag}\left(  d_{\eta}^{\alpha
}\right)  _{\alpha=1,\ldots,r}$, where $d_{\eta}^{\alpha}$, that we assume
positive, represents the growth rate of the population in patch $\alpha$
during a slow time interval. Regarding the fast process, we represent
migrations between patches under environment $\eta$ by a column-stochastic
primitive matrix $\mathbf{P}_{\eta}\in\mathbb{R}_{+}^{r\times r}$. We denote
$\boldsymbol{v}_{\eta}=\left(  v_{\eta}^{1},\ldots,v_{\eta}^{r}\right)  $ the
unique (positive) right eigenvector of matrix $\mathbf{P}_{\eta}$ associated
to eigenvalue 1 whose entries sum up to 1.

In this case model \eqref{modcomp2} reads
\begin{equation}
\boldsymbol{x}_{k}(t+1)=\mathbf{D}_{\tau_{t+1}}(\mathbf{P}_{\tau_{t+1}}%
)^{k}\boldsymbol{x}_{k}(t). \label{mcs1}%
\end{equation}

Regarding model \eqref{modcomp3}, let $s_{\eta}^{\alpha}>0$ be the survival
rate in patch $\alpha$ and environment $\eta$. Then we define $\tilde{d}%
_{\eta}^{\alpha}>0$ through $\tilde{d}_{\eta}^{\alpha}=s_{\eta}^{\alpha
}/d_{\eta}^{\alpha},$ and denote $\mathbf{\tilde{D}}_{\eta}=\text{diag}\left(
\tilde{d}_{\eta}^{\alpha}\right)  _{\alpha=1,\ldots,r}$ and $\mathbf{S}%
_{k,\eta}=\text{diag}\left(  (s_{\eta}^{\alpha})^{1/k}\right)  _{\alpha
=1,\ldots,r}$ to obtain
\begin{equation}
\tilde{\boldsymbol{x}}_{k}(t+1)=\mathbf{\tilde{D}}_{\tau_{t+1}}(\mathbf{S}%
_{k,\tau_{t+1}}\mathbf{P}_{\tau_{t+1}})^{k}\tilde{\boldsymbol{x}}_{k}(t).
\label{mcs2}%
\end{equation}

In both cases, the reduced system is a scalar system in which the only
variable is the total population size.

For system \eqref{mcs1} the reduced system is
\begin{equation}
y(t+1)=\hat{h}_{\tau_{t+1}}y(t), \label{res1}%
\end{equation}
with
\begin{equation}
\hat{h}_{\eta}=\sum_{\alpha=1}^{r}v_{\eta}^{\alpha}d_{\eta}^{\alpha},\ \eta
\in\mathcal{I}. \label{res11}%
\end{equation}

In the case of system \eqref{mcs2}, the reduced system becomes
\begin{equation}
\label{res2}\tilde{y}(t+1)=\tilde{h}_{\tau_{t+1}}\tilde{y}(t),
\end{equation}
where in this case
\begin{equation}
\tilde{h}_{\eta}=\exp\left(  \sum_{\alpha=1}^{r}v_{\eta}^{\alpha}\log(s_{\eta
}^{\alpha})\right)  \sum_{\alpha=1}^{r}v_{\eta}^{\alpha}\tilde{d}_{\eta
}^{\alpha},\ \eta\in\mathcal{I}. \label{res21}%
\end{equation}

Since (\ref{res1}) and \eqref{res2} are scalar, we are able to find analytical
expressions for their SGRs, $\hat{\lambda}_{\mathrm{S}}$ and $\tilde{\lambda
}_{\mathrm{S}}$, and their SLVs, $\hat{\sigma}^{2}$ and $\tilde{\sigma}^{2}$.

We recall that we assume condition C1, and so the Markov chain $\tau_{t}$ has
a stationary probability distribution $\boldsymbol{\pi}=(\pi_{1},\pi
_{2},...,\pi_{n})$. Since the $d_{\eta}^{\alpha}$ and the $\tilde{d}_{\eta
}^{\alpha}$ are positive and so are vectors $\boldsymbol{v}_{\eta}$, then the
$\hat{h}_{\eta}$ and $\tilde{h}_{\eta}$ are also positive and as a consequence
the sets $\left\{  \hat{h}_{1},\ldots,\hat{h}_{n}\right\}  $ and $\left\{
\tilde{h}_{1},\ldots,\tilde{h}_{n}\right\}  $ are ergodic, so Condition C2
holds. Thus, for large enough $k$ all four systems \eqref{mcs1}, \eqref{mcs2},
\eqref{res1} and \eqref{res2} meet the sufficient hypotheses for the existence
of an asymptotic lognormal distribution for population size.

Using (\ref{pepito}) we can obtain explicitly the SGR and the SLV that
characterize the asymptotic distribution for systems \eqref{res1} and
\eqref{res2}, the reason being that in these cases the normalized population
$y(t)/\left\Vert y(t)\right\Vert $ is trivial (equal to 1 with probability
one), and so the stationary distribution of the chain $(\tau_{t}%
,y(t)/\left\Vert y(t)\right\Vert )$ is simply that of $\tau_{t}$. Then it is
immediate to obtain that
\begin{align*}
\log\hat{\lambda}_{\mathrm{S}} &  =\overset{n}{\underset{\eta=1}{\sum}}%
\pi_{\eta}\log\hat{h}_{\eta}=\overset{n}{\underset{\eta=1}{\sum}}\pi_{\eta
}\log\left(  \sum_{\alpha=1}^{r}v_{\eta}^{\alpha}d_{\eta}^{\alpha}\right)  ,\\
\log\tilde{\lambda}_{\mathrm{S}} &  =\overset{n}{\underset{\eta=1}{\sum}}%
\pi_{\eta}\log\tilde{h}_{\eta}=\overset{n}{\underset{\eta=1}{\sum}}\pi_{\eta
}\left(  \sum_{\alpha=1}^{r}v_{\eta}^{\alpha}\log(s_{\eta}^{\alpha}%
)+\log\left(  \sum_{\alpha=1}^{r}v_{\eta}^{\alpha}\tilde{d}_{\eta}^{\alpha
}\right)  \right)  .
\end{align*}
Regarding the SLV, for the sake of simplicity in the mathematical expressions
we will only consider the IID case, i.e., the case with no serial correlation
in $\tau_{t}$ (the general expressions can be found in \cite{Alonso09}). In
that case it is immediate to conclude that
\begin{align*}
\hat{\sigma}^{2} &  =\overset{n}{\underset{\eta=1}{\sum}}\pi_{\eta}\left(
\log\hat{h}_{\eta}\right)  ^{2}-\left(  \overset{n}{\underset{\eta=1}{\sum}%
}\pi_{\eta}\log\hat{h}_{\eta}\right)  ^{2},\\
\tilde{\sigma}^{2} &  =\overset{n}{\underset{\eta=1}{\sum}}\pi_{\eta}\left(
\log\tilde{h}_{\eta}\right)  ^{2}-\left(  \overset{n}{\underset{\eta=1}{\sum}%
}\pi_{\eta}\log\tilde{h}_{\eta}\right)  ^{2}.
\end{align*}

\subsection{Two-stage models\label{sec32}}

We consider a population structured into two stage-specific classes:
non-reproductive juveniles (class 1) and reproductive adults (class 2). The
demography, which governs the transition between the different classes, is
defined by the survival rates of juveniles and adults, the maturation rates of
juveniles and the fertility rates of adults.

We represent the state of the population at time $t$ by vector%
\[
\boldsymbol{x}(t):=(\boldsymbol{x}^{1}(t),\boldsymbol{x}^{2}(t))^{\mathsf{T}%
}\in{\mathbb{R}}_{+}^{2r},
\]
where $\boldsymbol{x}^{i}(t):=(x^{i1}(t),\ldots,x^{ir}(t))\in\mathbb{R}%
_{+}^{r}$ and $x^{i\alpha}(t)$ denotes the population density of class $i$ in
patch $\alpha$.

Let $\alpha\in\{1,...,r\}$, $j\in\{1,2\}$ and $\eta\in\mathcal{I}$. For each
patch $\alpha$ and each environment $\eta,$ let $s_{\eta}^{j,\alpha}$ be the
fraction of individuals of class $j$ alive at time $n$ that survive to time
$n+1$. Also, let $m_{\eta}^{\alpha}$ be the fraction of the surviving
juveniles that mature and become adults. Finally, supposing that reproduction
happens at the end of each period of time $[t,t+1)$, let $f_{\eta}^{\alpha}$
be the number of juveniles produced by an adult individual in patch $\alpha$
that has survived to time $t+1$.

Let us define matrices
\[
\boldsymbol{S}_{\eta}^{i}:=\text{diag}\left(  s_{\eta}^{i,1},...,s_{\eta
}^{i,r}\right)  \text{, }\boldsymbol{M}_{\eta}:=\text{diag}\left(  m_{\eta
}^{1},...,m_{\eta}^{r}\right)  ,\ \boldsymbol{F}_{\eta}:=\text{diag}\left(
f_{\eta}^{1},...,f_{\eta}^{r}\right)  .
\]
Then, the projection matrix $\mathbf{D}_{\eta}$ corresponding to demography
for environment $\eta$ consistent with the ordering of variables in
$\mathrm{X}(t)$ is
\[
\mathbf{D}_{\eta}=\left(
\begin{array}
[c]{cc}%
\boldsymbol{S}_{\eta}^{1}\left(  \boldsymbol{I}-\boldsymbol{M}_{\eta}\right)
& \boldsymbol{S}_{\eta}^{2}\boldsymbol{F}_{\eta}\\
\boldsymbol{S}_{\eta}^{1}\boldsymbol{M}_{\eta} & \boldsymbol{S}_{\eta}^{2}%
\end{array}
\right)  \in\mathbb{R}^{2r\times2r}.
\]

Regarding the fast process, for each environment $\eta$ and each class $i$, we
represent migrations between patches by a column-stochastic primitive matrix
$\boldsymbol{P}_{\eta}^{i}\in\mathbb{R}_{+}^{r\times r}$. We denote
$\boldsymbol{v}_{\eta}^{i}=\left(  v_{\eta}^{i,1},\ldots,v_{\eta}%
^{i,r}\right)  $ the unique positive right eigenvector of matrix
$\boldsymbol{P}_{\eta}^{i}$ associated to eigenvalue 1 whose entries sum up to
1. The matrix representing migration for the whole population is
$\mathbf{P}_{\eta}=$diag$\left(  \boldsymbol{P}_{\eta}^{1},\boldsymbol{P}%
_{\eta}^{2}\right)  $.

Therefore, in this setting model (\ref{modcomp2}) has the form%
\begin{equation}
\boldsymbol{x}_{k}(t+1)=\mathbf{D}_{\tau_{t+1}}\left(  \mathbf{P}_{\tau_{t+1}%
}\right)  ^{k}\boldsymbol{x}_{k}(t). \label{2clasesI}%
\end{equation}
Following the procedure outlined in Section \ref{sec22}, its corresponding
reduced system is%
\begin{equation}
\boldsymbol{y}(t+1)=\boldsymbol{\hat{H}}_{\tau_{t+1}}\boldsymbol{y}(t),
\label{2clasesIagreg}%
\end{equation}
where%
\begin{equation}
\boldsymbol{\hat{H}}_{\eta}:=\left(
\begin{array}
[c]{cc}%
\overset{r}{\underset{\alpha=1}{\sum}}s_{\eta}^{1,\alpha}\left(  1-m_{\eta
}^{\alpha}\right)  v_{\eta}^{1,\alpha} & \overset{r}{\underset{\alpha=1}{\sum
}}s_{\eta}^{2,\alpha}\phi_{\eta}^{\alpha}v_{\eta}^{2,\alpha}\\
\overset{r}{\underset{\alpha=1}{\sum}}s_{\eta}^{1,\alpha}m_{\eta}^{\alpha
}v_{\eta}^{1,\alpha} & \overset{r}{\underset{\alpha=1}{\sum}}s_{\eta
}^{2,\alpha}v_{\eta}^{2,\alpha}%
\end{array}
\right)  \in\mathbb{R}^{2\times2}\ ,\ \eta\in\mathcal{I}. \label{coefI}%
\end{equation}

We now turn our attention to model (\ref{modcomp3}), i.e., the model in the
case we re-scale survival to the fast scale. In this case, following the
proposed technique we define%
\begin{align*}
\mathbf{\tilde{D}}_{\eta}  &  :=\left(
\begin{array}
[c]{cc}%
\boldsymbol{I}-\boldsymbol{M}_{\eta} & \boldsymbol{F}_{\eta}\\
\boldsymbol{M}_{\eta} & \boldsymbol{I}%
\end{array}
\right)  \in\mathbb{R}^{2r\times2r},\\
\boldsymbol{S}_{k,\eta}^{i}  &  :=\text{diag}\left(  \left(  s_{\eta}%
^{i,1}\right)  ^{1/k},...,\left(  s_{\eta}^{i,r}\right)  ^{1/k}\right)
\in\mathbb{R}^{r\times r},\ \mathbf{S}_{k,\eta}:=\text{diag}\left(
\boldsymbol{S}_{k,\eta}^{1},\boldsymbol{S}_{k,\eta}^{2}\right)  \in
\mathbb{R}^{2r\times2r},
\end{align*}
and the resulting system has the form%
\begin{equation}
\mathrm{\tilde{x}}_{k}(t+1)=\mathbf{\tilde{D}}_{\tau_{t+1}}\left(
\mathbf{S}_{k,\tau_{t+1}}\,\mathbf{P}_{\tau_{t+1}}\right)  ^{k}\mathrm{\tilde
{x}}_{k}(t). \label{2clasesII}%
\end{equation}

To obtain the corresponding reduced system we define $\gamma_{\eta}^{i}%
:=\exp\left(  \overset{r}{\underset{\alpha=1}{\sum}}v_{\eta}^{i,\alpha}\log
s_{\eta}^{i,\alpha}\right)  $ and then in this setting system (\ref{modagreg3}%
) takes the form
\begin{equation}
\boldsymbol{\tilde{y}}(t+1)=\boldsymbol{\tilde{H}}_{\tau_{t+1}}%
\boldsymbol{\tilde{y}}(t), \label{2clasesIIagreg}%
\end{equation}
where%
\begin{equation}
\boldsymbol{\tilde{H}}_{\eta}:=\left(
\begin{array}
[c]{cc}%
\gamma_{\eta}^{1}\overset{r}{\underset{\alpha=1}{\sum}}\left(  1-m_{\eta
}^{\alpha}\right)  v_{\eta}^{1,\alpha} & \gamma_{\eta}^{2}\overset
{r}{\underset{\alpha=1}{\sum}}\phi_{\eta}^{\alpha}v_{\eta}^{2,\alpha}\\
\gamma_{\eta}^{1}\overset{r}{\underset{\alpha=1}{\sum}}m_{\eta}^{\alpha
}v_{\eta}^{1,\alpha} & \gamma_{\eta}^{2}%
\end{array}
\right)  \in\mathbb{R}^{2\times2}\ ,\ \eta\in\mathcal{I}. \label{coefII}%
\end{equation}

The reduced systems (\ref{2clasesIagreg}) and (\ref{2clasesIIagreg}) are not
scalar and so in general they are not analytically tractable. There are,
however, several particular cases in which analytical calculations are
possible. Here we focus on one of such cases.

Let us suppose that the parameters that govern demography are independent of
the environment (and therefore in the notation we drop subindex $\eta$) and
that migration rates depend on the environment through a common multiplicative
random variable, i.e., different environments increase or decrease all
migration rates by the same factor. Specifically, for each $i=1,2$ and
$\alpha\neq\beta$ we have $p_{\eta}^{i,\alpha\beta}=\xi_{\eta}^{i}%
t^{i,\alpha\beta}$ where $\xi_{\eta}^{i}\in(0,1]$ and the $t^{i,\alpha\beta
}\geq0$ verify that $\sum_{\beta\neq\alpha}t^{i,\alpha\beta}\leq1$. For
$\beta=\alpha$ we have $p_{\eta}^{i,\alpha\alpha}=1-\xi_{\eta}^{i}\sum
_{\beta\neq\alpha}t^{i,\alpha\beta}$. Then it is easy to check that the
eigenvector $\boldsymbol{v}_{\eta}^{i}$ of $\boldsymbol{P}_{\eta}^{i}$
associated to $1$ is independent of $\eta$, i.e., $\boldsymbol{v}_{\eta}%
^{i}=\boldsymbol{v}^{i}=\left(  v^{i,1},...,v^{i,r}\right)  ^{\text{T}}$ for
all $\eta\in\mathcal{I}$. So, we have that, although systems (\ref{2clasesI})
and (\ref{2clasesII}) are stochastic, the reduced systems (\ref{2clasesIagreg}%
) and (\ref{2clasesIIagreg}) are deterministic. More specifically we have
$\boldsymbol{\hat{H}}_{\eta}=\boldsymbol{\hat{H}}$,$\ \boldsymbol{\tilde{H}%
}_{\eta}=\boldsymbol{\tilde{H}}$ where $\boldsymbol{\hat{H}}$ and
$\boldsymbol{\tilde{H}}$ are given by (\ref{coefI}) and (\ref{coefII}) by
dropping subindex $\eta$. Therefore, $\hat{\lambda}_{\mathrm{S}}$ and
$\tilde{\lambda}_{\mathrm{S}}$ are, respectively, the dominant eigenvalues of
matrices $\boldsymbol{\hat{H}}$ and $\boldsymbol{\tilde{H}}$ whereas
$\hat{\sigma}^{2}=\tilde{\sigma}^{2}=0$.

\subsection{Comparison of the two modelling approaches}

\label{sec33}

In this section we use a very simple setting to illustrate the differences
between models \eqref{modcomp2} and \eqref{modcomp3}. Due to the large number
of parameters involved in the general case, we restrict our attention to the
case of Section \ref{sec31} of a population without structure and, further, we
assume a deterministic setting and that the environment is constituted by two
patches. We present a first particular case that yields the same growth rates
for both systems. Then, a second situation in which the re-scaled system
always possesses a smaller growth rate. And, finally, a third case where,
depending on parameter values, the growth rate of any of the two systems can
be larger than the other. In the latter, we show in particular that using one
of the models can predict exponential growth whereas, for the same parameter
values, the use of the other predicts extinction.

Following the notation in Section \ref{sec31}, let the demographic parameters
of the model be $d^{1}=\tilde{d}^{1}s^{1}$ and $d^{2}=\tilde{d}^{2}s^{2}$
where $\tilde{d}^{1}$ and $\tilde{d}^{2}$ denote the fertility coefficients in
each patch. The fast process is then represented by matrix
\[
\boldsymbol{P}=\left(
\begin{array}
[c]{cc}%
1-m_{2} & m_{1}\\
m_{2} & 1-m_{1}%
\end{array}
\right)  ,
\]
where $m_{1}\in(0,1)$ represents the migration rate from patch 2 to patch 1
and $m_{2}\in(0,1)$ the one from patch 1 to patch 2. $\boldsymbol{P}$ is a
primitive stochastic matrix with associated stable probability distribution
vector $\boldsymbol{v}=(v^{1},v^{2})=\left(  m_{1}/(m_{1}+m_{2}),m_{2}/\left(
m_{1}+m_{2}\right)  \right)  $. In this case model \eqref{modcomp2} reads:
\begin{equation}
\boldsymbol{x}_{k}(t+1)=\left(
\begin{array}
[c]{cc}%
\tilde{d}^{1}s^{1} & 0\\
0 & \tilde{d}^{2}s^{2}%
\end{array}
\right)  \left(
\begin{array}
[c]{cc}%
1-m_{2} & m_{1}\\
m_{2} & 1-m_{1}%
\end{array}
\right)  ^{k}\boldsymbol{x}_{k}(t), \label{mc2-33}%
\end{equation}
and model \eqref{modcomp3}
\begin{equation}
\tilde{\boldsymbol{x}}_{k}(t+1)=\left(
\begin{array}
[c]{cc}%
\tilde{d}^{1} & 0\\
0 & \tilde{d}^{2}%
\end{array}
\right)  \left(
\begin{array}
[c]{cc}%
(s^{1})^{\frac{1}{k}}(1-m_{2}) & (s^{1})^{\frac{1}{k}}m_{1}\\
(s^{2})^{\frac{1}{k}}m_{2} & (s^{2})^{\frac{1}{k}}(1-m_{1})
\end{array}
\right)  ^{k}\tilde{\boldsymbol{x}}_{k}(t). \label{mc3-33}%
\end{equation}
Their corresponding reduced systems are, respectively,
\[
y(t+1)=\hat{h}\,y(t)=\left(  v^{1}\tilde{d}^{1}s^{1}+v^{2}\tilde{d}^{2}%
s^{2}\right)  y(t),
\]
and
\[
\tilde{y}(t+1)=\tilde{h}\,\tilde{y}(t)=(s^{1})^{v^{1}}(s^{2})^{v^{2}}\left(
v^{1}\tilde{d}^{1}+v^{2}\tilde{d}^{2}\right)  \tilde{y}(t).
\]
Thus, we have to compare $\hat{h}=v^{1}\tilde{d}^{1}s^{1}+v^{2}\tilde{d}%
^{2}s^{2}$ and $\tilde{h}=(s^{1})^{v^{1}}(s^{2})^{v^{2}}\left(  v^{1}\tilde
{d}^{1}+v^{2}\tilde{d}^{2}\right)  $.

\noindent\textbf{Case 1:} Assuming equal survival rates, $s^{1}=s^{2}$, we
have
\[
\tilde{h}=s^{v^{1}}s^{v^{2}}\left(  v^{1}\tilde{d}^{1}+v^{2}\tilde{d}%
^{2}\right)  =s\left(  v^{1}\tilde{d}^{1}+v^{2}\tilde{d}^{2}\right)  =\hat
{h}.
\]
There is no difference between the asymptotic growth rates of systems
\eqref{mc2-33} and \eqref{mc3-33}.

\noindent\textbf{Case 2:} Assuming equal fertility rates, $\tilde{d}%
^{1}=\tilde{d}^{2}=\tilde{d}$ and using the inequality relating the (weighted)
arithmetic mean and the (weighted) geometric mean, we obtain
\[
\tilde{h}=s^{v^{1}}s^{v^{2}}\left(  v^{1}\tilde{d}+v^{2}\tilde{d}\right)
=s^{v^{1}}s^{v^{2}}\tilde{d}\leq\left(  v^{1}s^{1}+v^{2}s^{2}\right)
\tilde{d}=\hat{h},
\]
i.e., the asymptotic growth rate of the re-scaled model \eqref{mc3-33} is always
smaller than the one of system \eqref{mc2-33}.

We would like to stress that the first two cases can be straightforwardly
extended to an environment with an arbitrary number of patches.

\noindent\textbf{Case 3:} Assuming a uniform distribution of individuals
between patches, $v^{1}=v^{2}=1/2$, we can find, depending on parameters
values, the same results as in cases 1 and 2 and, in addition, the reverse
inequality $\hat{h}<\tilde{h}$. The expressions for $\hat{h}$ and $\tilde{h}$
in this particular case are
\[
\hat{h}=\frac{1}{2}(\tilde{d}^{1}s^{1}+\tilde{d}^{2}s^{2})\ \text{ and
}\ \tilde{h}=\frac{1}{2}\sqrt{s^{1}s^{2}}(\tilde{d}^{1}+\tilde{d}^{2}).
\]
Considering $\tilde{d}^{1}$ and $\tilde{d}^{2}$ constant, we can write the
fraction of the survival rates in both patches, $C:=\hat{h}/\tilde{h}$, as a
function of variable $\alpha:=s^{2}/s^{1}\in(0,\infty)$ in the following way:
\[
C(\alpha)=\frac{1}{\sqrt{\alpha}}\frac{\tilde{d}^{1}}{\tilde{d}^{1}+\tilde
{d}^{2}}+\sqrt{\alpha}\frac{\tilde{d}^{2}}{\tilde{d}^{1}+\tilde{d}^{2}}.
\]
An elementary analysis yields that $C$ is decreasing in $(0,\alpha_{min})$ and
increasing in $(\alpha_{min},\infty)$, with $\alpha_{min}=\tilde{d}^{1}%
/\tilde{d}^{2}$ and
\[
C(\alpha_{min})=\frac{\sqrt{\tilde{d}^{1}\tilde{d}^{2}}}{(\tilde{d}^{1}%
+\tilde{d}^{2})/2}\leq1.
\]
On the other hand, for $\tilde{d}^{1}\neq\tilde{d}^{2}$, equation
$C(\alpha)=1$ has two roots, $\alpha_{1}=1$ and $\alpha_{2}=(\tilde{d}%
^{1}/\tilde{d}^{2})^{2}$, one in $(0,\alpha_{min})$ and the other in
$(\alpha_{min},\infty)$.

We can conclude from the above that, if $\tilde{d}^{1}<\tilde{d}^{2},$ then
$\hat{h}<\tilde{h}$ for $s^{2}/s^{1}\in\left(  (\tilde{d}^{1}/\tilde{d}%
^{2})^{2},1\right)  $ and $\hat{h}>\tilde{h}$ for $s^{2}/s^{1}\in\left(
0,(\tilde{d}^{1}/\tilde{d}^{2})^{2}\right)  \cup(1,\infty)$. An analogous
reverse result is obtained in the case $\tilde{d}^{1}>\tilde{d}^{2}$.

An important additional question is whether $\hat{h}$ and $\tilde{h}$ can be
one larger than 1 and the other less than 1, since this would mean that one of
the models predicts extinction whereas the other predicts exponential growth.
The answer is positive. Let us fabricate an example to illustrate this fact.
Take, for instance, the previous case with $\tilde{d}^{1}<\tilde{d}^{2}$ and
$s^{2}/s^{1}\in\left(  (\tilde{d}^{1}/\tilde{d}^{2})^{2},1\right)  ,$ what
implies $\hat{h}<\tilde{h}$. Now just change parameters $\tilde{d}^{1}$ and
$\tilde{d}^{2}$ into $2\tilde{d}^{1}/\left(  \hat{h}+\tilde{h}\right)  $ and
$2\tilde{d}^{2}/\left(  \hat{h}+\tilde{h}\right)  $. It is immediate to see
that the corresponding growth rates, that we call $\hat{h}^{\prime}$ and
$\tilde{h}^{\prime}$, verify the required condition:
\[
\hat{h}^{\prime}=\dfrac{2}{\hat{h}+\tilde{h}}\hat{h}<1<\dfrac{2}{\hat
{h}+\tilde{h}}\tilde{h}=\tilde{h}^{\prime}.
\]

The previous discussion corresponds to a deterministic setting. If we
introduce stochasticity, the number of parameters is highly increased, what
renders the model difficult to analyze. However, one can check that the three
aforementioned cases still hold.

\section{Discussion}

\label{sec4} We have dealt with the issue of distinguishing time scales in
discrete systems to help in their analysis. The setting is, on the one hand,
simple because we deal with a linear structured metapopulation model and, on
the other hand, complex due to the fact that the model considers environmental
stochasticity and admits an arbitrary numbers of individual classes and
spatial patches.

When there are two processes acting at different time scales that must be
gathered in a single discrete model, it is reasonable to choose the slow time
unit to express it. In this way, the action of the fast process can be
represented by letting it act a number of times approximately equal to the
ratio between the two time scales. It is not easy to establish a criterion to
classify processes between those that occur at the slow time scale and those
happening at the fast time scale. Indeed, let us consider processes, such as
mortality, predation and others, which are often measured at the slow time
scale. If they act almost continuously, then it can be argued that they should
be better considered as occurring at the fast time scale.

We have analysed the particular case of survival in a linear stochastic
structured metapopulation model. First, we have proposed a model in which
survival is included in the slow process. In a second step, we have shown how
to express its action on a much shorter interval of time by performing a sort
of $k$-th root of their action in a slow time unit. Using this we have
proposed a second model, to be compared with the first one, in which survival
has been re-scaled to the fast time scale.

The comparison of the two models is done through their SGRs and SLVs, as they
are the main parameters describing their asymptotic behaviour. In general, it
is not possible to calculate them exactly, and so the available course of
action is to estimate them by means of computer simulations or other
approximations. The obtained reduction results simplify this task by
performing it for the associated reduced models, what is much less costly.

Considering the particular case of unstructured populations, their reduced
models are scalar, what allows one to obtain closed expressions for their SGRs
and SLVs, that approximate the SGRs and the SLVs of the original models.

In this setting of unstructured populations, we carry out a comparison of the
two models and we present cases where the re-scaling of survival makes no
difference in the dynamics, other cases where considering survival at the fast
scale reduces population growth rate and, finally, some cases where the
population growth can be larger in any of the two models depending on
parameters values.

\section{Conclusion}

\label{sec5}

We have shown the relevance of using time scales when modelling through
dynamical systems. This is done in the framework of linear matrix models with
environmental stochasticity.

A key point to build an accurate two time scale model is the choice of the
time scale that it is associated to each of the processes involved. For
a given process, data can be obtained on a certain time scale and, at the same
time, they are better included in the model on a different time scale. To deal
with this issue, we have considered the procedure of re-scaling applied to the
survival process. Survival data are translated from the slow to the fast time
scale. We have proposed some cases where including survival at either the slow
or the fast time scale can mean the difference between exponential growth or
extinction of the population. This illustrates how important the appropriate choice of
time scale can be.

Directly linked to the time scales models are the methods to reduce them. They
are the tools to simplify their analysis. In \ref{app2} we have presented an
extension of an already existing reduction technique for linear matrix models
with environmental stochasticity. This new result has lead to reducing the
model with re-scaled survival, but it can be apply in more general situations.

The re-scaling in the case proposed in this work is rather straightforward,
but this is not always the case and it would be very interesting to study the
influence of time scale choice on other demographic processes. \cite{Nguyen11}
proposed a more general approach to re-scaling in a
deterministic setting that can be extended to the stochastic case along the guidelines
and the reduction results presented here.

\appendix

\section{Matrix models with environmental stochasticity}

\label{app1} This section presents the basic form of the matrix models that
consider environmental stochasticity when the environmental process is a
Markov chain. We assume that the population lives in an ambient in which there
are different environmental states that, for simplicity, we suppose finite and
label with $\eta\in\mathcal{I}:=\left\{  1,...,n\right\}  $; a nonnegative
matrix $\boldsymbol{A}_{\eta}\in\mathbb{R}^{N\times N}$ represents the vital
rates of the population in environment $\eta$. The environmental variation is
characterized by a Markov chain $\tau_t$, $t=0,1,2,...$ defined on a certain
probability space $(\Omega,\mathcal{F},p)$ over the state space $\mathcal{I}$.
For each realization $\omega\in\Omega$ of the process, the population is
subject to environmental conditions $\tau_{t+1}(\omega)$ between times $t$ and
$t+1$. Thus, the model reads
\begin{equation}
\boldsymbol{z}(t+1)=A_{\tau_{t+1}}\boldsymbol{z}(t), \label{e1}%
\end{equation}
where $\boldsymbol{z}(t)=\left(  z^{1}(t),...,z^{N}(t)\right)  ^{\mathsf{T}}$
represents the population vector at time $t$ for each $t=0,1,...$ . We assume
that $\boldsymbol{z}_{0}$ is a fixed nonzero nonnegative vector.

The following theorem gives important information on the distribution of the
total population size $\left\Vert \boldsymbol{z}(t)\right\Vert _{1}=\left\vert
z^{1}(t)\right\vert +\cdots+\left\vert z^{N}(t)\right\vert $ (in the sequel
the subscript in the norm will be omitted):

\begin{theorem}
\label{fust} Let us assume that Markov chain $\tau_{t}$\ is homogeneous,
irreducible and aperiodic and, further, that the set $\mathcal{A}%
=\{\boldsymbol{A}_{\eta},\ \eta=1,...,n\}$\ of vital rate matrices is ergodic,
i.e., there exists a positive integer $g$ such that any product of $g$
matrices (with repetitions allowed) drawn from $\mathcal{A}$ is a positive
matrix (i.e., its components are all positive). Then we have:

1. We can define the stochastic growth rate (SGR) $\lambda_{\mathrm{S}}$ for
system (\ref{e1}) through $\log\lambda_{\mathrm{S}}:=\underset{t\rightarrow
\infty}{\lim}\log\left\Vert \boldsymbol{z}(t)\right\Vert /t$ with probability
one. Moreover, $\lambda_{\mathrm{S}}$ is finite, and is independent of the
initial probabilities of the chain states and of the initial (nonzero)
population vector $\boldsymbol{z}_{0}\geq\boldsymbol{0}$ and can be calculated
through
\begin{equation}
\log\lambda_{\mathrm{S}}=\mathbb{E}_{F}\log\left\Vert \boldsymbol{A}%
_{\tau_{t+1}}\frac{\boldsymbol{z}(t)}{\left\Vert \boldsymbol{z}(t)\right\Vert
}\right\Vert , \label{pepito}%
\end{equation}
where $F$ denotes the stationary distribution for $\left(  \tau_{t}%
,\boldsymbol{z}(t)/\left\Vert \boldsymbol{z}(t)\right\Vert \right)  $ (whose
existence is guaranteed \citep{Cohen77}).

2. We can also define the \textquotedblleft\textit{scaled logarithmic
variance}\textquotedblright\ (SLV) as $\sigma^{2}:=\lim_{t\rightarrow\infty
}\mathbb{V}\left[  \log\left\Vert \boldsymbol{z}(t)\right\Vert \right]  /t$
where $\mathbb{V}$ denotes variance and where $\sigma^{2}$, which is finite,
is independent of the initial probabilities of the chain and of the initial
(nonzero) population vector $\boldsymbol{z}_{0}\geq\boldsymbol{0}$.

3. If $\sigma^{2}>0$ the population size is asymptotically lognormal in the
sense that
\[
\frac{\log\left\Vert \boldsymbol{z}(t)\right\Vert -t\log\lambda_{\mathrm{S}}%
}{t^{1/2}\sigma}\overset{\mathcal{L}}{\longrightarrow}\mathcal{N}\left(
0,1\right)
\]
where $\mathcal{N}\left(  0,1\right)  $ denotes a normal distribution of zero
mean and unit variance and $\mathcal{L}$ denotes convergence in distribution.
\end{theorem}

\begin{proof}The result is essentially Lemma 4 in \cite{Tulja80}
except for a technical detail. See Theorem 2.3. in \cite{Alonso09} for the proof.
\end{proof}

\section{Reduction of matrix models with environmental stochasticity}

\label{app2}

We present here a general class of matrix models with environmental
stochasticity which can be reduced, as well as the reduction technique. We
also include a result that relates the behavior of the original and the
reduced model. These models and results generalize those of \cite{Alonso09},
which are only valid for models of the kind (\ref{modcomp2}), to more general
models like (\ref{modcomp3}). For the sake of simplicity, in this work we deal
with a finite number of environments, but the general technique is
valid even when there is an infinite (denumerable or not) number of environments.

Let $N,\,n\in\mathbb{N}$ and $\mathcal{I}=\left\{  1,...,n\right\}  $. For
each $k$ large enough, we consider a set of nonnegative matrices
$\mathcal{A}_{k}=\{\boldsymbol{H}_{k,\eta},\ \eta\in\mathcal{I}\}$ with
$\boldsymbol{H}_{k,\eta}\in\mathbb{R}_{+}^{N\times N}$ and denote the
population vector $\boldsymbol{x}_{k}(t)=(x_{k}^{1}(t),\ldots,x_{k}^{N}%
(t))\in\mathbb{R}^{N}$. Let $\tau_{t}$ be the Markov chain defined in Appendix A that selects the environment in each time step. Then we define the so
called \textit{complete model}
\begin{equation}
\boldsymbol{x}_{k}(t+1)=\boldsymbol{H}_{k,\tau_{t+1}}\boldsymbol{x}_{k}(t),
\label{modcomp}%
\end{equation}
where we assume that $\boldsymbol{x}_{k}(0)$ is a fixed nonzero vector
$\boldsymbol{x}_{0}\geq\boldsymbol{0}$.

In order to reduce system (\ref{modcomp}), we suppose that $k$ is large enough
and we impose some conditions which are specified in the following two hypotheses:

\begin{hypothesis}
\label{HA} For all $\eta\in\mathcal{I}$ there exists a matrix
$\boldsymbol{\bar{H}}_{\eta}$ such that
\[
\lim_{k\longrightarrow\infty}\boldsymbol{H}_{k,\eta}=\boldsymbol{\bar{H}%
}_{\eta}.
\]

\end{hypothesis}

\begin{hypothesis}
\label{HB} There exist $q\in\mathbb{N}$, $q<N$, such that for all $\eta
\in\mathcal{I}$ we can decompose $\boldsymbol{\bar{H}}_{\eta}$ in the form
\begin{equation}
\label{dfr}\boldsymbol{\bar{H}}_{\eta}=\boldsymbol{D}_{\eta}\boldsymbol{G},
\end{equation}
where $\boldsymbol{G}\in\mathbb{R}_{+}^{q\times N}$, independent of
environment $\eta$, and $\boldsymbol{D}_{\eta}\in\mathbb{R}_{+}^{N\times q}$
are nonnegative matrices.
\end{hypothesis}

In what follows we accept Hypotheses \ref{HA} and \ref{HB}. Then, we proceed
to reduce system (\ref{modcomp}) in two steps.

First, we define the so-called \textit{auxiliary system} which approximates
(\ref{modcomp}) when $k\rightarrow\infty$. Denoting its vector of variables at
time $t$ by $\boldsymbol{x}(t)$, this auxiliary system reads
\begin{equation}
\boldsymbol{x}(t+1)=\boldsymbol{\bar{H}}_{\tau_{t+1}}\boldsymbol{x}%
(t)=\boldsymbol{D}_{\tau_{t+1}}\boldsymbol{Gx}(t). \label{modaux}%
\end{equation}
The set of environmental matrices for the auxiliary system is $\mathcal{A}%
_{aux}=\left\{  \boldsymbol{\bar{H}}_{\eta},\ \ \eta\in\mathcal{I}\right\}  $.

Now we define \textit{global variables}, which will play the role of state
variables of the reduced, or \emph{aggregated}, system
\begin{equation}
\boldsymbol{y}(t):=\boldsymbol{Gx}(t)\in\mathbb{R}_{+}^{q}. \label{varglob}%
\end{equation}

Multiplying both sides of (\ref{modaux}) with $\boldsymbol{G}$, we obtain the
aggregated system
\[
\boldsymbol{y}(t+1)=\boldsymbol{Gx}(t+1)=\boldsymbol{GD}_{\tau_{t+1}%
}\boldsymbol{Gx}(t)=\boldsymbol{GD}_{\tau_{t+1}}\boldsymbol{y}(t),
\]
which is a stochastic system for the global variables $\boldsymbol{y}(t)$ that
we use as an approximation of system (\ref{modcomp}).

Denoting
\begin{equation}
\boldsymbol{\hat{H}}_{\eta}:=\boldsymbol{GD}_{\eta}\ ,\ \ \eta\in\mathcal{I},
\label{g23}%
\end{equation}
we can write the aggregated system as
\begin{equation}
\boldsymbol{y}(t+1)=\boldsymbol{\hat{H}}_{\tau_{t+1}}\boldsymbol{y}(t),
\label{modagreg}%
\end{equation}
for which the set of environmental matrices is $\mathcal{A}_{ag}=\left\{
\boldsymbol{\hat{H}}_{\eta},\ \eta\in\mathcal{I}\right\}  .$

Note that through the previous procedure we have constructed an approximation
of (\ref{modcomp}) for $k$ large enough that allows us to reduce a system with
$N$ variables to a new system with $q$ variables. In most practical
applications \citep{Caswell01,Rogers15}, $q$ will be much smaller than $N$.

In order to obtain asymptotic results of the original system \eqref{modcomp}
from the aggregated system \eqref{modagreg} we proceed to relate the
distribution of the total population size for both systems. For the results to
hold we need to impose the following hypothesis:

\begin{hypothesis}
\label{H20} For all $\eta\in\mathcal{I},$ matrices $\boldsymbol{D}_{\eta}$ are
row-allowable (i.e., each row has at least a nonzero component) and matrix
$\boldsymbol{G}$ is column-allowable (i.e., each column has at least a nonzero component).
\end{hypothesis}

\begin{proposition}
\label{prop500} Assume that the environmental process $\tau_{t}$ is an
irreducible and aperiodic homogeneous Markov chain and that Hypotheses
\ref{HA}, \ref{HB} and \ref{H20} hold. Then, if $\mathcal{A}_{ag}$ is ergodic,
the set $\mathcal{A}_{aux}$ is also ergodic and there exists $k_{0}%
\in\mathbb{N}$ such that $\mathcal{A}_{k}$ is ergodic for all $k\geq k_{0}$.
Therefore, if $\mathcal{A}_{ag}$ is ergodic, the aggregated system meets the
sufficient conditions of Theorem \ref{fust} for the SGR and SLV to exist, and
those sufficient conditions are also met by the auxiliary system and by the
original system for $k\geq k_{0}$.
\end{proposition}

\begin{proof} A straightforward generalization of the proofs of Proposition 4.1.
and Corollary 4.2. in \cite{Alonso09}.
\end{proof}

Therefore, if $\mathcal{A}_{ag}$\ is an ergodic set, we have, under the
hypotheses of Proposition \ref{prop500}, that we can define the following
characteristics for the

\vspace{2ex}

\noindent Aggregated system: \ $\log\hat{\lambda}_{\mathrm{S}}:=\underset
{t\rightarrow\infty}{\lim}\dfrac{\mathbb{E}\log\left\Vert \boldsymbol{y}%
(t)\right\Vert }{t}$ \ ; \ $\hat{\sigma}^{2}:=\lim_{t\rightarrow\infty}%
\dfrac{1}{t}\mathbb{V}\left[  \log\left\Vert \boldsymbol{y}(t)\right\Vert
\right]  $;

\vspace{2ex}

\noindent Original system: \ \ $\log\lambda_{\mathrm{S},k}:=\underset
{t\rightarrow\infty}{\lim}\dfrac{\mathbb{E}\log\left\Vert \boldsymbol{x}%
_{k}(t)\right\Vert }{t}$ \ ; \ $\sigma_{k}^{2}:=\lim_{t\rightarrow\infty
}\dfrac{1}{t}\mathbb{V}\left[  \log\left\Vert \boldsymbol{x}_{k}(t)\right\Vert
\right]  $.

\vspace{2ex}

Next theorem shows that, when $k$ is large, we can approximate $\lambda_{\mathrm{S},k}$
and $\sigma_{k}$ through $\hat{\lambda}_{\mathrm{S}}$ and $\hat{\sigma}$, respectively,
which are easier to compute.

\begin{theorem}
\label{th11}Let us assume the same Hypotheses of Proposition \ref{prop500}
and, in addition, that $\mathcal{A}_{ag}$\ is an ergodic set. Then we have%
\[
\underset{k\rightarrow\infty}{\lim}\lambda_{\mathrm{S},k}=\hat{\lambda
}_{\mathrm{S}},\ \lim_{k\rightarrow\infty}\sigma_{k}=\hat{\sigma},
\]
and so the lognormal asymptotic distribution for the original system can be
approximated through the parameters $\hat{\lambda}_{\mathrm{S}}$ and
$\hat{\sigma}$ that correspond to the reduced system.
\end{theorem}

\begin{proof} A quite technical generalization of the proof of Theorem 4.3. in
\cite{Alonso09} that will appear published elsewhere.
\end{proof}

\section*{Acknowledgements}

Authors are supported by Ministerio de Econom\'{\i}a y Competitividad (Spain),
Project MTM2014-56022-C2-1-P.

\bibliography{ECOMOD2020}

\end{document}